# Temperature, Pressure, Velocity, and Water Vapor Mole Fraction Profiles in a Ramjet Combustor using Dual Frequency Comb Spectroscopy and a High Temperature Absorption Database

David Yun[a,*], Scott C. Egbert[a], Nathan A. Malarich[a], Ryan K. Cole[a], Jacob J. France[b], Jiwen Liu[c], Kristin M. Rice[d], Mark A. Hagenmaier[d], Jeffrey M. Donbar[d], Nazanin Hoghooghi[a], Sean C. Coburn[a], Gregory B. Rieker[a,*]

[a]*Precision Laser Diagnostics Laboratory, University of Colorado Boulder, Boulder, CO 80309, USA*
[b]*Innovative Scientific Solutions Incorporated, Dayton, OH 45459, USA*
[c]*Taitech Incorporated, Beavercreek, OH 45430, USA*
[d]*U.S. Air Force Research Laboratory, Wright-Patterson AFB, OH 45433, USA*
*Corresponding authors: david.yun@colorado.edu, greg.rieker@colorado.edu*

**Abstract:** Accurate diagnostics of the combustor region of ramjet engines can improve engine design and create benchmarks for computational fluid dynamics models. Previous works demonstrate that dual frequency comb spectroscopy can provide low uncertainty diagnostics of multiple flow parameters in the non-combusting regions of ramjets. However, the high temperatures present in the combustor present a challenge for broadband spectroscopic absorption models that are used to interpret measurements in these regions. Here, we utilize a new water vapor absorption database created for high temperature water-air mixtures to fit spectra measured in a ground-test ramjet engine with a broadband near-infrared dual comb absorption spectrometer. We extract 2D profiles of pressure, temperature, water mole fraction, and velocity using this new database. We demonstrate that the new database provides the lowest fit residuals compared to other water vapor absorption databases. We compare computational fluid dynamics simulations of the combustor with the measured data to demonstrate that the simulations overpredict heat release and water vapor production.

## 1. Introduction

Ramjet engines fly through the atmosphere at supersonic speeds and have the potential for efficient, long-distance flight, and reusability. Ramjets ingest air through an inlet and compress it through a series of oblique shocks that form from the interaction of the air with the narrowing inlet geometry. Though the shocks slow down the air, the air still enters the combustor at high speed, with combustion occurring at slightly subsonic speeds for ramjets. Thus, a primary challenge to ramjet flight is maintaining stable combustion despite the high air speed through the engine. Studying critical design parameters for enabling stable flight requires diagnostic methods that provide accurate information in the harsh conditions of the combustor.

Many studies have explored how different combustor design factors can promote stable combustion such as injection schemes [1–9], combustor geometry [10–16], and equivalence ratio [17–19]. Diagnostic methods can characterize the effect of these design factors on combustion by measuring parameters such as temperature, pressure, velocity, and species mole fractions. Physical sensors such as pressure transducers, thermocouples and thermocouple rakes [1,3,4,9,10,12,13,15–18,20–23] are commonly used because they are reliable, have low uncertainties, and are relatively easy to implement. However, when these sensors are used for measurements in the core flow away from the wall they can be intrusive and cause flow distortion. On the other hand, optical techniques can provide measurements of flow conditions less intrusively and are commonly used in hypersonic combustor research. Shadowgraph [17,21,24] and schlieren imaging [5,22,19,23] show density gradients such as shock locations inside the combustor. Particle image velocimetry [25–27] maps the movement of the general flow. Planar laser induced fluorescence [9,28] and chemiluminescence [17,19,29] use laser- or reaction-induced emissions to target the location of certain molecules or chemical reactions inside the combustor; and scattering techniques such as coherent anti-Stokes Raman scattering [30,31] are used to measure temperatures and species mole fractions within the flow. While all of these optical techniques have their own unique advantages, they are typically limited to measuring one or two thermofluidic parameters and may require complex optical experimental configurations. Laser absorption spectroscopy (LAS) is an optical technique that can simultaneously measure multiple flow parameters (temperature, pressure, species concentration, and velocity) and can be implemented with relatively simple optics and optical access. Several studies have used tunable diode LAS in ramjet combustors [32–37].

Another form of LAS called dual frequency comb spectroscopy (DCS) has recently been demonstrated in a ramjet for the first time [38]. DCS is capable of reaching low measurement uncertainties by covering a broad spectral range (capturing many absorption features simultaneously) with high spectral resolution and extremely accurate, stable, and precise optical frequency referencing and control. A follow-on to ref. [38] demonstrated DCS for spatially resolved velocity measurements with less than 5% uncertainty in a combustor [39]. In that same work, spatially resolved low-uncertainty measurements of other flow quantities (temperature, pressure, water concentration, and air mass flux) are demonstrated for the non-combusting region of the ramjet. Flow parameters other than velocity were not reported for the combustor due to the elevated measurement uncertainties resulting from high-temperature error in the spectral databases available for fitting the measurements. The previous best-option for a spectroscopic database consisted of high-temperature pure water absorption parameters measured by Schroeder et al. [40] with additional parameters calculated by Antony et al. [41], but lacked the accurate high-temperature air-broadening values needed for accurately retrieving temperature, pressure, and species mole fraction in the combustor. This limitation did not impact combustor velocity retrievals since the velocimetry used a dual-beam configuration which only relies on Doppler shift of the measured absorption and therefore minimizes the effect of database uncertainty on velocimetry [38].

In this work, we revisit the measurements of [39] but now employ a new database that was created to model high-temperature water-air mixtures as described in Egbert et al. [42,43]. Utilizing the new database, we accurately derive temperature, pressure, water mole fraction, and velocity across a 2D profile in the combustor. We compare fit residuals from this database to those using other databases - HITRAN2020 [44] and the aforementioned hybrid database from Schroeder et al. [40] and Antony et al. [41] (which hereafter we refer to as S&A). Finally, we use our results generated using the new database to evaluate combustor CFD.

## 2. Spectroscopic Model Databases

In LAS, a laser is passed through a gas sample and the attenuation of the laser light as a function of wavelength is measured. The attenuation is due to absorption of the light at wavelengths that are resonant with quantum transitions of the molecules in the sample. The total absorption as a function of wavelength, $\alpha(\lambda)$, for a particular molecular species is the summation of absorption transitions ($j$) as defined by Beer's Law in Eq. 1:

$$\alpha(\lambda) = P\chi_s L \sum_j S_j(T, E'')\phi_j(\lambda_{0j}, \lambda, T, P, U, \chi_s, \chi_{f1}, \dots, \chi_{fn}) \quad (1)$$

Here, $P$ is pressure, $T$ is temperature, $U$ is bulk velocity, $\chi_s$ is mole fraction of the absorbing molecule (subscript "s" denotes "self"), $\chi_f$ is the mole fraction of other molecules in the gas mixture (subscript "f" denotes "foreign"), $S_j$ is the linestrength of transition j, $E''$ is the lower state energy of the transition, $\phi_j$ is the lineshape function, and $\lambda_{0j}$ is the linecenter. As shown in Beer's law, absorption is dependent on sample conditions (temperature, pressure, species mole fraction, etc.) which ultimately enables retrieving these parameters from absorption measurements. Some sample conditions affect all absorption lines uniformly while others impact different features uniquely. For instance, bulk velocity applies a uniform Doppler shift to all line positions. Similarly, pressure, pathlength, and mole fraction of the absorbing molecule uniformly increase/decrease the overall absorption of all transitions. Conversely, the strength of a given transition has a unique temperature dependence based on its lower state energy. The width of the lineshape broadens with increasing partial pressures of its collisional partners and is defined by unique broadening parameters for both self- and foreign-collisional partners. Each line will also experience a pressure dependent linecenter shift that is defined by self- and foreign-shift parameters.

These unique parameters for each transition of a molecule are found in spectroscopic databases and are used to create models for evaluating measured spectra to determine sample conditions. Thus, the accuracy of a measurement using absorption spectroscopy relies on the accuracy of the spectroscopic database underpinning the parameter retrievals. One of the more widely used spectroscopic databases is HITRAN, whose most recent iteration is HITRAN2020 [44]. HITRAN is used for a variety of applications from exoplanetary and terrestrial studies [45–47] to combustion research [48–50]. In previous DCS measurements in this ramjet, we used a high-temperature optimized water vapor absorption database from Schroeder et al. [40] which derived linestrengths, linecenters, self-broadening and self-shift parameters [51] from laboratory measurements of pure $H_2O$ vapor for the wavelength region used in the current work. Since the laboratory measurements were only of pure water vapor, air broadening parameters are taken from HITRAN2012 [51]. We also supplemented the database with self-broadening temperature-dependent parameters from Antony et al. [41]. Here, we refer to this combined database as S&A. While this database exhibited low fit residuals for lower temperature measurements in the non-combusting regions of the ramjet (<900 K) [38,39], there were still large fit residuals when attempting to retrieve parameters from the high temperature regions of the combustor (up to 2000 K). Therefore temperature, pressure, and species concentrations were not reported for the combustor.

In the current work, we fit ramjet combustor data using HITRAN2020, S&A, and a new measured database from Egbert et al. [42,43]. The Egbert database utilized high-temperature pure-water *and* water-air laboratory spectra to derive not only the aforementioned constants in S&A as well as high-temperature air-broadening and air-shift parameters for over 5000 features. Additionally, over 500 new high-temperature water features were identified that are missing from the previous databases. Given the low concentration of water, these updates to air-broadened widths and shifts are particularly relevant in the hot flows of the combustor where air is the dominant collisional partner for water molecules. Figure 1 demonstrates the difference between these three databases for absorption models derived from each database at conditions experienced in the combustor (T = 1800 K, P = 0.95 atm, $\chi_{H2O}$ = 14%). The differences in broadening, strength, and position of transitions between these models will result in different parameter retrievals when using the different models to fit the same measured spectra.

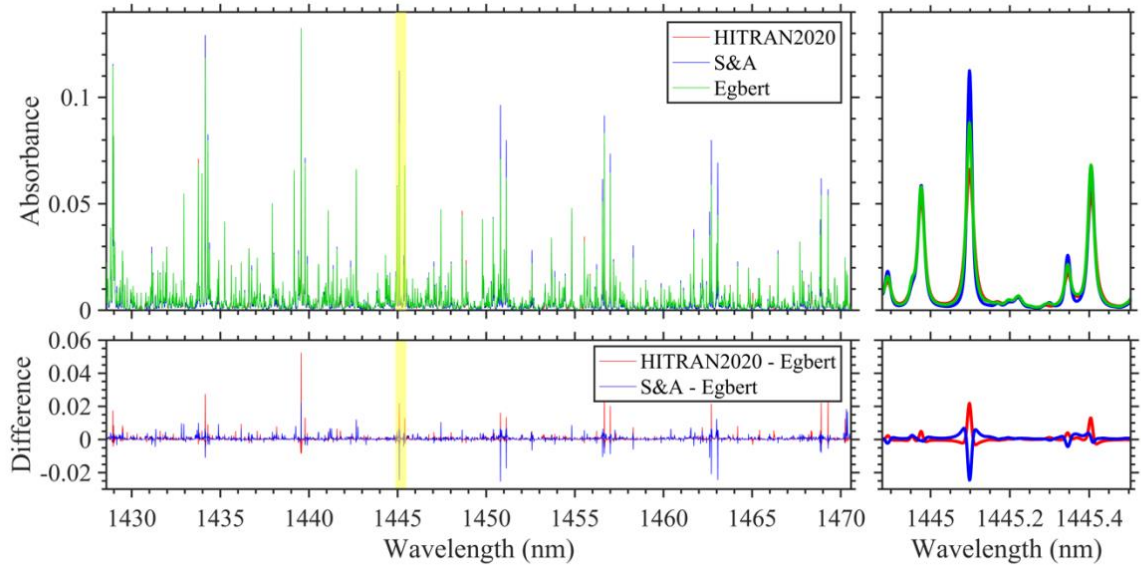

*Figure 1. Comparison of spectroscopic models derived from the different databases at conditions of 1800 K, 0.95 atm, 14% water vapor mole fraction, and 1100 m/s. The top plots show the generated models from the three databases: HITRAN2020 (red), S&A (blue), and Egbert (green). The left panel shows the full wavelength range of the measurements used in the current study and on the right is a zoomed-in view corresponding to the yellow highlighted region in the full spectra. The bottom plots show the difference between the models produced by HITRAN2020 and Egbert (red) and between S&A and Egbert (blue) with the full wavelength range on the left and a zoomed-in view on the right.*

### 3. Dual comb spectroscopy experiment

The experimental setup for these measurements is described briefly here, and in more detail in sections 2 and 4 of [39]. Measurements are taken with a near-infrared dual comb spectrometer (DCS) through the combustor of a grounded, dual-mode ramjet test engine at Wright Patterson Air Force Base. The DCS is based on fully stabilized mode-locked frequency comb lasers which produce light over a large wavelength range and consist of discretely and narrowly spaced wavelengths of light called comb "teeth" [52]. We tune the combs to best capture high-temperature water absorption transitions from 1430 – 1470 nm (6800 – 7000 cm$^{-1}$). The combs have a ~200 MHz pulse repetition rate which produce comb "teeth" with a fine spectral spacing of ~1.4 pm (0.0067 cm$^{-1}$). The position and spacing of the teeth are locked using a common continuous-wave, narrow-linewidth reference laser and a *f-2f* phase locking scheme [53–57]. The DCS control electronics time base is linked to a GPS-disciplined oscillator. These control measures result in a comb tooth spacing accuracy of 25 ppb [58]. In DCS, two combs are tuned to have slightly different repetition rates (626 Hz in this case). The different repetition rates result in different comb tooth spacing such that when the combs are combined on a photodetector, interference signals arise at radio frequencies that are directly proportional to the optical frequencies contained in the laser spectrum but can be easily detected by a fast photodetector [59].

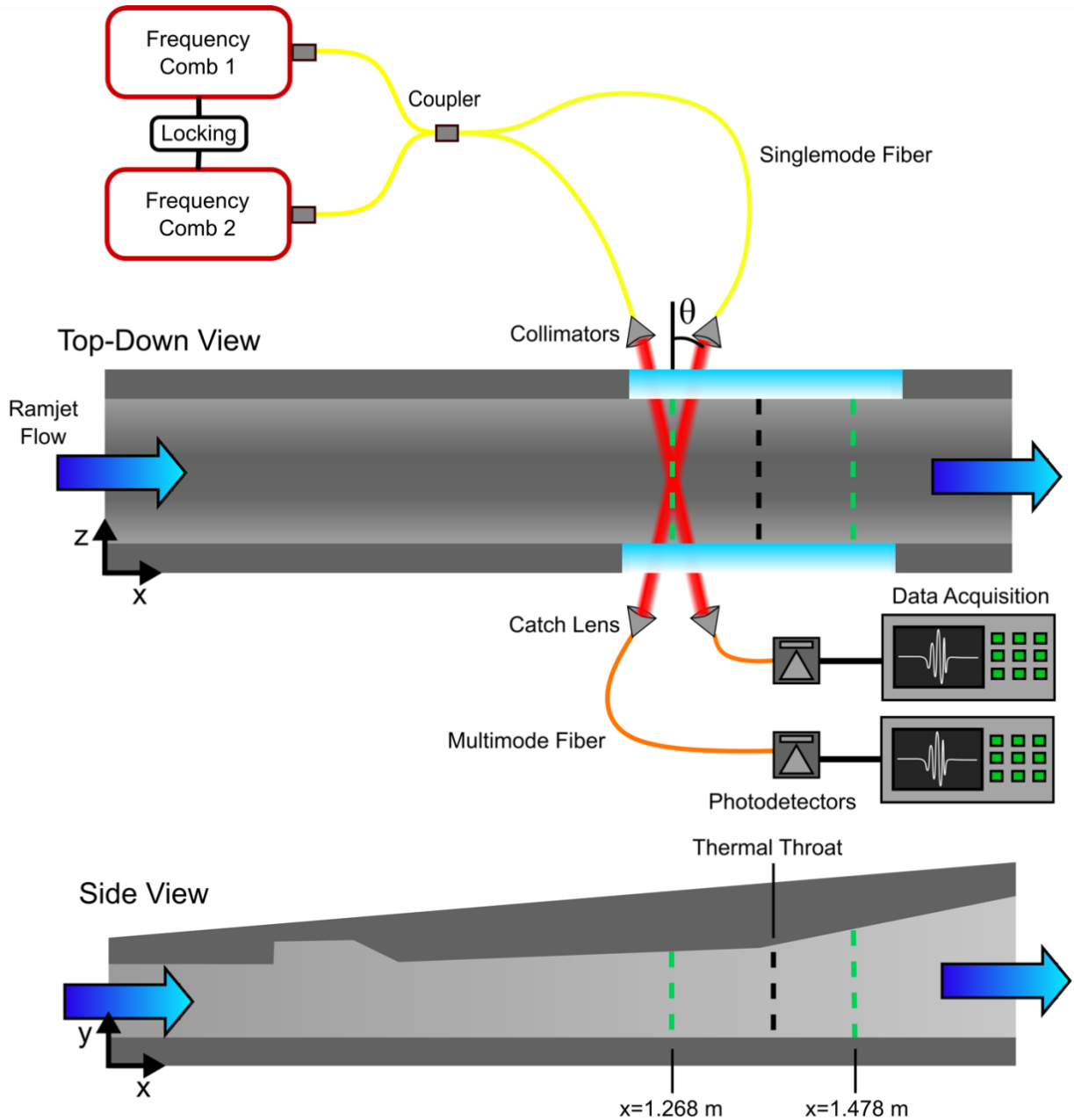

*Figure 2. Schematic of experimental setup. Light from two frequency combs is combined in a coupler (yellow). The combined light travels down two fiber channels to the combustor. Collimators send the light through the combustor via quartz optical access windows in a crossed-beam configuration parallel to the floor of the combustor and angled along the axial (x) axis as shown in the top-down view. Catch lenses focus the light onto multimode fibers (orange) which send the light to photodetectors. Data acquisition devices read the signals from the photodetectors. We translate the optics to take two scans along the height (y) of the combustor at axial locations, x=1268 mm and x=1478 mm as shown in the side view.*

Figure 2 contains a schematic of the experimental configuration for the DCS measurements in the combustor. Light from the two frequency combs is combined and passed through the combustor in a crossed-beam configuration (Fig. 2). The two channels of light are parallel to the floor of the combustor and are at equal but opposite angles ($\pm 15°$) relative to the axial axis (along the flow direction). We measure light from each channel on separate photodetectors and record the measurement signals on separate data acquisition devices. The measurement location is varied along both the axial and vertical directions via two motorized stages to obtain two vertical profiles of the combustor, each with 14 heights, taken at positions 1268 mm and 1478 mm after the combustor entrance. These locations are before and after the thermal throat, respectively, which is where the subsonic flow in the combustor is accelerated back to supersonic speeds. As described in [39], measurements at each location are averaged for approximately 45 seconds by combining data from a series of steady 5-second duration combustion events.

## 4. Data reduction

We fit the measured spectra to models created from the three spectroscopic databases. For each measurement, spectra from both crossed-beam channels are fit simultaneously for a common temperature, pressure, velocity, and water mole fraction. We use a Levenberg-Marquadt non-linear least-squares minimization fitting method that employs cepstral analysis [60]. This method fits the signal in cepstral space which is the inverse Fourier transform of the absorbance signal and is a powerful technique as it is effectively immune to laser power baseline effects. Figure 3 compares fits using the aforementioned three databases: HITRAN2020 [44], S&A [40,41], and the new Egbert database [42,43].

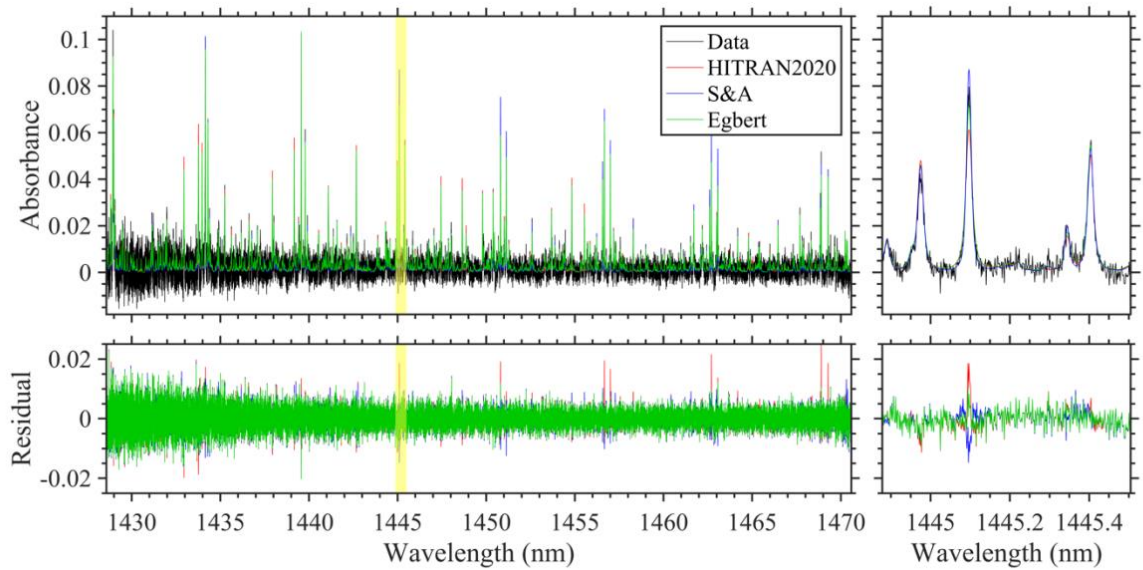

*Figure 3. DCS data (black) from the downstream-propagating channel at measurement location x=1478 mm, y=42 mm with fit using three different spectroscopic databases: HITRAN2020 (red), S&A (blue), and Egbert et al. (green). The plots on the left show the fits across the whole measurement spectral range (1430 – 1470 nm). The top plot shows the DCS data with the fitted models from the three databases overlaid and the bottom plot shows the residuals from the fits. The plots on the right show the same information but zoomed in to a narrower spectral range corresponding to the transparent yellow region (1444.9 – 1445.5 nm).*

We see substantial differences between the fits using the three databases. The HITRAN2020 fit demonstrates residual structures associated with linestrength and broadening errors. The S&A fit residuals are lower than those of HITRAN2020 and mostly attributable to broadening errors. Egbert has the lowest residuals of all three with structures primarily indicating small broadening differences. We note that none of the databases include the influence of $CO_2$ on broadening, and thus do not expect to achieve zero residual when there is up to several percent $CO_2$ present.

The best test for the databases would be under controlled, well-known laboratory conditions, however these are challenging to recreate for the ground-test ramjet. Instead, we compare the overall quality of these fits using the reduced chi-square statistic which describes the contribution of the model-data mismatch to the residual. A reduced chi-square of 1 indicates a residual that is due only to ~0.004 absorbance noise in the data, whereas a reduced chi-square above 1 indicates a contribution of model-data mismatch to the residual. The top and middle panels of Fig. 4 show the reduced chi-square value for each measurement fitted with each database. The bottom panel contains boxplots to summarize this data. For each individual measurement, the trend between databases for lowest to highest reduced chi-square is consistently Egbert, S&A, and then HITRAN2020. The boxplots in the bottom panel of Fig. 4 show that the reduced chi-square mean for Egbert is lower than for the other two. These results suggest that Egbert produces models that fit consistently better to the combustor data. Better fits with reduced residual typically indicate improved database accuracy.

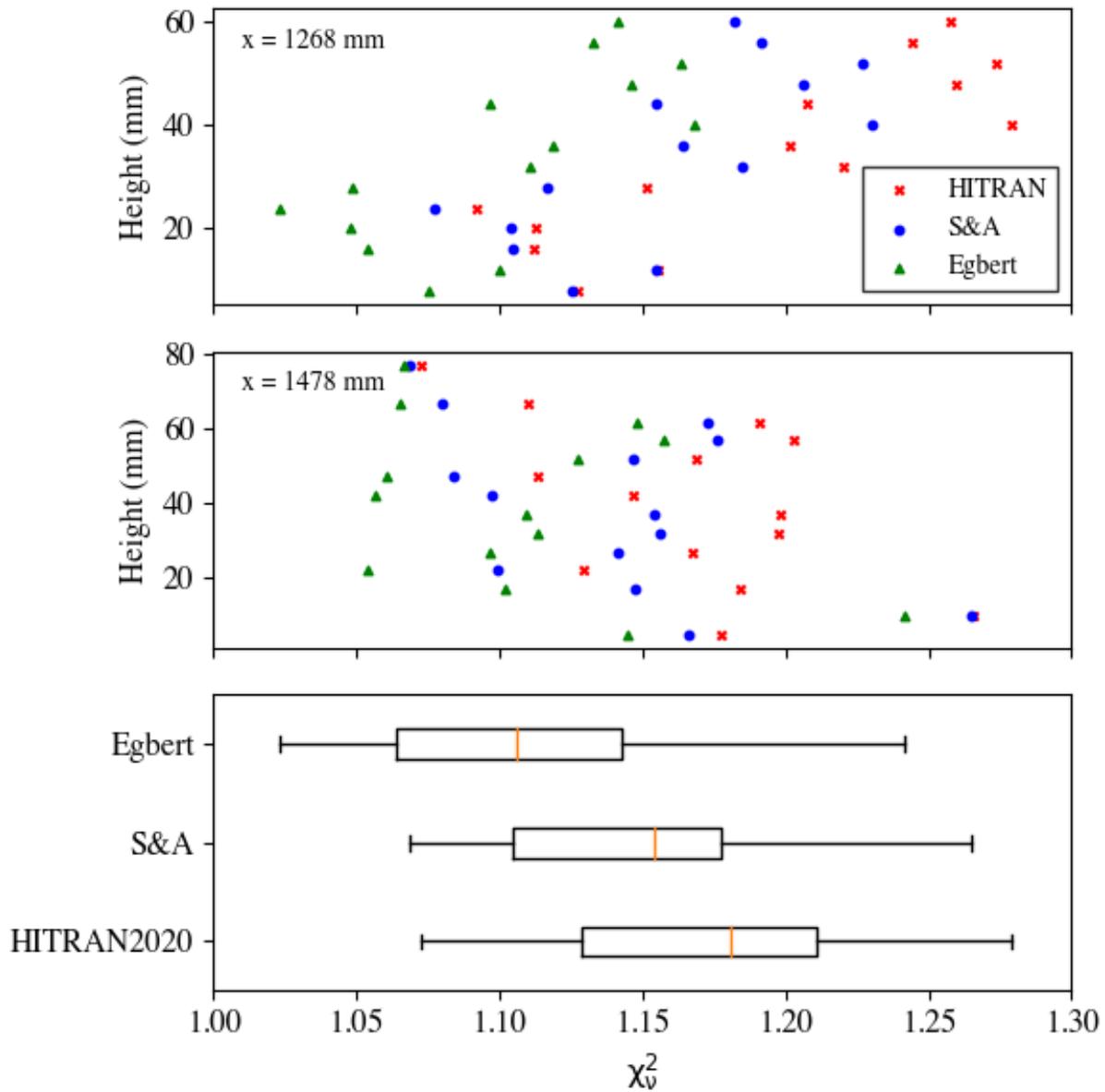

*Figure 4. Plots of the reduced chi-square statistic for combustor data fits using the three databases - HITRAN2020 (red x), S&A (blue circle), and for Egbert (green triangle). The top plot demonstrates the individual chi-square statistics at different heights at the x = 1268 mm axial location and the middle plot shows reduced chi-square statistics for the axial location x = 1478 mm. The bottom plot shows boxplots for each database across all measurements at both axial locations.*

In Fig. 5 we demonstrate vertical profiles of velocity, pressure, temperature, and water mole fraction for both axial locations. We show the retrieved parameters using all three databases. For velocity, all databases effectively give the same values – this consistency is to be expected as a two beam setup allows the velocity measurement to be minimally influenced by database error [38]. For pressure, all databases show the same general trend, however the S&A and Egbert results match closely with each other while HITRAN2020 is ~25% lower. For temperature, all databases also fit the same general trend where temperature increases with height. However, HITRAN2020 fits lower temperatures overall, S&A higher temperatures, and Egbert in between the two. Trends in water mole fraction are not as consistent as the other parameters – particularly at the x = 1478 mm location which shows more scatter due to lower absorption (note the 2x decrease in pressure between the two axial locations in the ramjet). However, the measurements do show that HITRAN2020 returns the highest mole fractions, while Egbert shows slightly higher mole fractions relative to S&A at lower heights and slightly lower mole fractions at higher heights.

We use bold green markers for the parameters retrieved with Egbert to denote that we recommend these parameters as the most accurate and use them for comparison with the computational fluid dynamics. We indicate calculated uncertainty for the Egbert retrieved parameters using the same method described in [39] where we consider uncertainty contributions from the instrument frequency axis, beam angle, noise precision, and background subtraction. Here, we do not add the database uncertainty to the calculated uncertainty as it is difficult to replicate these high temperatures in a controlled laboratory setting to validate against. The average uncertainties are 3.9%, 1.3%, 2.0%, and 3.9% for velocity, pressure, temperature, and mole fraction respectively.

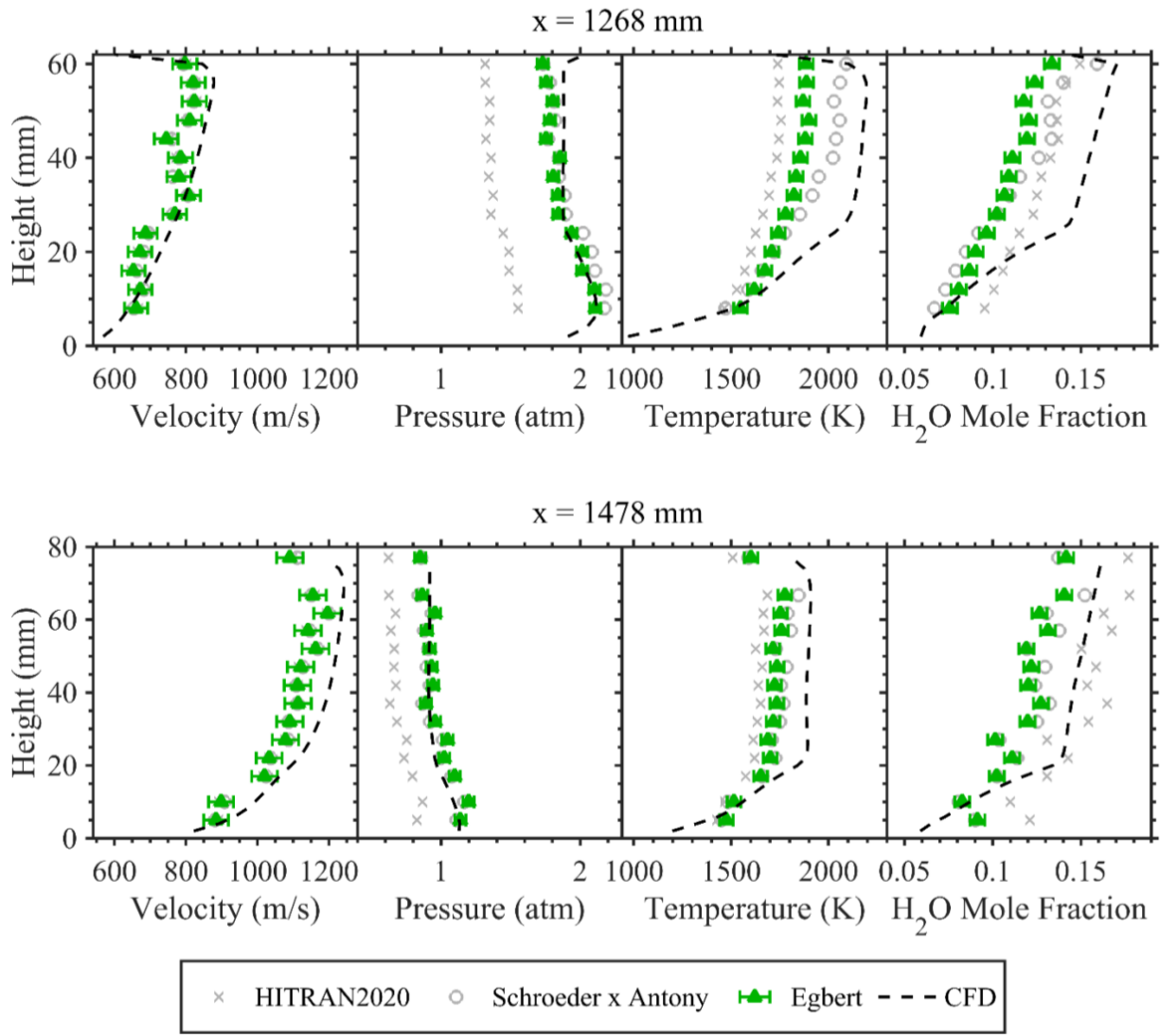

*Figure 5. Fit results by height for velocity, pressure, water ($H_2O$) mole fraction, and temperature. The top row shows fit results at the first axial location of x= 1268mm and the bottom row shows results for the second axial location of x = 1478 mm. Values are shown for fits with HITRAN2020 (grey x marker), S&A (grey circle marker), and Egbert (green triangle marker). We use bold green markers for the parameters retrieved with Egbert as we recommend these parameters as the most accurate. Error bars indicate calculated uncertainty. Values derived from CFD solution are over laid on top (black dashed line).*

In Fig. 5, we also include computational fluid dynamic (CFD) simulation-derived values. A full 3D CFD solution is performed for this work. Modeling the combustor flow is difficult due to the complex chemical reactions and compressible flow physics. Key modeling selections of the CFD were the use of the two-equation cubic $k - \epsilon$ turbulence model, and a reduced reaction mechanism for JP-7 [61] based on HyChem [62,63]. In addition, a calibration study varying the turbulent Schmidt number was performed to find the value that provides best agreement with the wall pressure distribution as measured with sidewall pressure taps. For the current work, a Schmidt number of 0.8 was selected.

We generate the CFD values in Fig. 5 using a process that enables direct comparison with the laser measurements. For each measurement location, we simulate spectra over discrete length bins along the line-of-sight of the laser beams for that measurement location through the 3D CFD parameter field. We then sum the individual spectra to create a path-integrated spectrum that we fit with the same algorithm as the DCS measurements. This technique accounts for any nonuniformity biases in the CFD in the same manner that they may influence the DCS measurements.

Figure 5 shows that the CFD values for pressure and velocity match closely with the laser measurements. The agreement between the CFD-derived and laser-measured path-integrated pressure values serve as a point of validation for the laser measurements because the CFD is calibrated to independent pressure readings with sidewall pressure taps. Prior to this study, it was not possible to benchmark the temperature and mole fraction in the CFD. Figure 5 shows that the CFD-derived values for both are significantly higher than the laser measurements. The higher temperature and water vapor mole fractions could indicate that the reaction mechanism predicts faster heat release and reaction completion (which would result in higher water vapor production and temperature) than what is actually occurring in the combustor, and serves as pathway for future CFD refinement.

## 5. Conclusion

Dual frequency comb spectroscopy continues to evolve as an important diagnostic method for ramjets, expanding here to include the challenging environment of the combustor. As with other absorption spectroscopy methods, the absorption database that informs spectral retrievals can be a major contributor to the diagnostic uncertainty. In this work, we demonstrate DCS measurements of temperature, pressure, water mole fraction and velocity across a 2D profile of a combustor flow and investigate how three different spectroscopic databases affect these results. These databases are the commonly used HITRAN2020 [44], a prior high-temperature database that contains self-collisional water vapor parameters based on Schroeder et al. [40] and Antony et al. [41] (S&A), and a new database that expands the bandwidth and includes high-temperature air-water parameters from Egbert et al. [42,43]. The resulting spread in results across databases for parameter retrieval at the high-temperature conditions of the combustor (except velocity, which is database independent) demonstrate the importance of databases on uncertainty. An assessment of the fit residuals and chi-square statistics indicate that the Egbert database facilitates the best fits for the combustor data, and is recommended to be the most accurate for these conditions. With the proper spectroscopic database, we demonstrate that the DCS technique can play an important role in understanding and improving models for high temperature flows. We compare CFD-derived parameter values to the profiles retrieved using the Egbert database. We show good agreement to the velocity and pressure values, the latter having been driven in the CFD by wall pressure measurements. The comparisons suggest that the CFD currently overpredicts the temperature and water creation in the combustor, and form the basis for future work to refine the CFD simulations.

## Acknowledgements

We would like to thank the test operators at RC-18, Steve Enneking, Andrew Baron, and Justin Stewart, who made sure we had everything we needed to take these measurements. This work has been cleared by the Air Force under case number AFRL-2023-4205.

## Funding

This research was sponsored by the Defense Advanced Research Projects Agency (W31P4Q-15-1-0011), Air Force Research Laboratory (FA8650-20-2-2418) and the Air Force Office of Scientific Research (FA9550-17-1-0224, FA8650-20-2-2418).